\documentclass[10pt,conference]{IEEEtran}

\usepackage[utf8]{inputenc}
\usepackage{booktabs}
\usepackage{subcaption}
\usepackage{amsmath}
\usepackage{amssymb}
\usepackage{textcomp}
\usepackage{xcolor}
\usepackage{pgfpages}
\usepackage{pgfplots}
\usepackage{pgfplotstable}
\usepackage{tikz}
\usepackage{enumerate}
\usepackage[noend]{algpseudocode}
\usepackage[linesnumbered,ruled,vlined]{algorithm2e}
\usepackage{listings}
\usepackage{tabularx}
\usepackage{url}
\usepackage[bottom]{footmisc}
\usepackage[font=bf]{caption}
\usepackage{multirow}
\usepackage{arydshln}
\usepackage{xspace}
\usepackage{flushend}
\usepackage{diagbox}

\usepackage{bm}
\usepackage{siunitx}
\usepackage{multirow}
\usepackage{enumitem}
\usepackage{tcolorbox}
\usepackage{datapie}
\usepackage[hidelinks]{hyperref}

\def\BibTeX{{\rm B\kern-.05em{\sc i\kern-.025em b}\kern-.08em
    T\kern-.1667em\lower.7ex\hbox{E}\kern-.125emX}}

\definecolor{light-gray}{gray}{0.75}
\definecolor{BrickRed}{rgb}{0.8, 0.25, 0.33}
\definecolor{lightorange}{HTML}{FFB74D}
\definecolor{Crimson}{rgb}{0.86, 0.08, 0.24}
\definecolor{NavyBlue}{rgb}{0.0, 0.0, 0.5}
\definecolor{DarkOrange}{rgb}{1.0, 0.55, 0.0}
\definecolor{Green}{rgb}{0.0, 0.5, 0.0}
\definecolor{Red}{rgb}{1.0, 0.0, 0.0}
\definecolor{Blue}{rgb}{0.0, 0.0, 1.0}
\definecolor{Yellow}{rgb}{1.0, 1.0, 0.0}
\definecolor{Orange}{rgb}{1.0, 0.65, 0.0}
\definecolor{Purple}{rgb}{0.62, 0.0, 0.77}
\definecolor{LimeGreen}{rgb}{0.2, 0.8, 0.2}
\definecolor{Cyan}{rgb}{0.0, 1.0, 1.0}
\definecolor{ForestGreen}{rgb}{0.13, 0.55, 0.13}

\makeatletter
\newenvironment{btHighlight}[1][]
{\begingroup\tikzset{bt@Highlight@par/.style={#1}}\begin{lrbox}{\@tempboxa}}
{\end{lrbox}\bt@HL@box[bt@Highlight@par]{\@tempboxa}\endgroup}

\newcommand\btHL[1][]{%
  \begin{btHighlight}[#1]\bgroup\aftergroup\bt@HL@endenv%
}
\def\bt@HL@endenv{%
  \end{btHighlight}%
  \egroup
}
\newcommand{\bt@HL@box}[2][]{%
  \tikz[#1]{%
    \pgfpathrectangle{\pgfpoint{0.3pt}{0pt}}{\pgfpoint{\wd #2}{\ht #2}}%
    \pgfusepath{use as bounding box}%
    \node[anchor=base west,fill=lightorange,outer sep=0pt,inner xsep=0.3pt,inner ysep=0pt,minimum height=\ht\strutbox+0.3pt,#1]{\raisebox{0.3pt}{\strut}\strut\usebox{#2}};
  }%
}
\makeatother

\usetikzlibrary{calc,trees,positioning,arrows,chains,shapes.geometric,%
  decorations.pathreplacing,decorations.pathmorphing,decorations.text,shapes,%
  matrix,shapes.symbols,patterns,shadows}

\pgfplotsset{width=7cm,compat=newest}
\usepgfplotslibrary{
  colorbrewer,
  groupplots
}

\algrenewcommand\alglinenumber[1]{\tiny\color{Black!70}{#1}}
\algrenewcommand\algorithmicforall[2]{\textbf{for} $i=$ #1 \textbf{to} #2}
\algrenewtext{ForAll}{\algorithmicforall}
\algnewcommand\algorithmicswitch{\textbf{switch}}
\algnewcommand\algorithmiccase{\textbf{case}}
\algdef{SE}[SWITCH]{Switch}{EndSwitch}[1]{\algorithmicswitch\ #1\ \algorithmicdo}{\algorithmicend\ \algorithmicswitch}%
\algdef{SE}[CASE]{Case}{EndCase}[1]{\algorithmiccase\ #1}{\algorithmicend\ \algorithmiccase}%
\algtext*{EndSwitch}%
\algtext*{EndCase}%
\algdef{SE}[SUBALG]{Indent}{EndIndent}{}{\algorithmicend\ }%
\algtext*{Indent}
\algtext*{EndIndent}

\lstdefinestyle{basic}{%
  morekeywords     = [1]{var},%
  morekeywords     = [2]{assert, assume},%
  keywordstyle     = \bfseries\color{DarkBlue},%
  keywordstyle     = [2]\bfseries\color{BrickRed},
  commentstyle     = \ttfamily\color{Black!70}\lst@ifdisplaystyle\footnotesize\fi,%
  basicstyle       = \ttfamily\lst@ifdisplaystyle\footnotesize\fi,%
  columns          = [c]fixed,%
  aboveskip        = 0mm,%
  belowskip        = 2mm,%
  keepspaces       = true,%
  mathescape       = true,%
  escapechar       = ?,%
  tabsize          = 2,%
  numbers          = left,%
  numberstyle      = \tiny\color{Black!80},%
  numbersep        = 1.0em,%
  stepnumber       = 1,%
  firstnumber      = 1,%
  showstringspaces = false,%
  captionpos       = b,%
  extendedchars    = true,%
  xleftmargin      = 2.5em,%
  upquote          = true,%
  abovecaptionskip = 0.5em,%
  belowcaptionskip = 0.5em,%
  moredelim        = **[is][{\btHL[fill=light-gray]}]{°}{°},%
}

\lstdefinestyle{clang}{%
  language         = C,%
  style            = basic,%
}
\renewcommand{\lstlistingname}{Listing}

\renewcommand{\figurename}{Figure}
\renewcommand{\tablename}{Table}
\renewcommand{\thetable}{\arabic{table}}

\newcommand\secref[1]{Sect.~\ref{#1}}
\newcommand\secsref[1]{Sects.~\ref{#1}}
\newcommand\figref[1]{Fig.~\ref{#1}}
\newcommand\figsref[1]{Figs.~\ref{#1}}
\newcommand\algoref[1]{Alg.~\ref{#1}}
\newcommand\tabref[1]{Tab.~\ref{#1}}

\newcommand\tool{\textsc{tAIlor}\xspace}
\newcommand\crab{\textsc{Crab}\xspace}

\newcommand{\narrowings}{\texttt{NUM\_\-NARROW\_\-ITERATIONS}\xspace}
\newcommand{\thresholds}{\texttt{NUM\_\-WIDEN\_\-THRESHOLDS}\xspace}
\newcommand{\delays}{\texttt{NUM\_\-DELAY\_\-WIDEN}\xspace}

\makeatletter
\newif\if@anonymize

\@anonymizetrue

\if@anonymize
  \newcommand\anonymize[1]{Link removed for double-blind review.}
\else
  \newcommand\anonymize[1]{\tiny#1}
\fi
\makeatother

\newcommand{\doub}[2]{\mathrm{{#1}_{#2}}}
\newcommand{\cur}[1]{\doub{#1}{curr}}
\newcommand{\best}[1]{\doub{#1}{best}}
\newcommand{\nex}[1]{\doub{#1}{next}}
\newcommand{\init}[1]{\doub{#1}{init}}
\newcommand{\mut}[1]{\doub{#1}{mut}}
\newcommand{\prog}{P}
\newcommand{\rmax}{r_{max}}
\newcommand{\lmax}{l_{max}}
\newcommand{\idom}{i_{dom}}
\newcommand{\iset}{i_{set}}
\newcommand{\cost}{cost}

\newcommand{\rec}{\mathrm{rec}}

\newcommand{\act}{\mathrm{act}}
\newcommand{\add}{\mathrm{ADD}}
\newcommand{\modfy}{\mathrm{MOD}}
\newcommand{\new}[1]{\mathrm{#1}_{\mathrm{new}}}
\newcommand{\modact}{\mathrm{act}_{\mathrm{m}}}
\newcommand{\ingr}{\mathrm{ingr}}
\newcommand{\pgt}{\mathrm{GT}}
\newcommand{\plt}{\mathrm{LT}}
\newcommand{\inc}{\mathrm{INC}}

\DTLsetpiesegmentcolor{1}{Red!60}
\DTLsetpiesegmentcolor{2}{Green!60}
\DTLsetpiesegmentcolor{3}{Blue!60}
\DTLsetpiesegmentcolor{4}{Yellow!60}
\DTLsetpiesegmentcolor{5}{Orange!60}
\DTLsetpiesegmentcolor{6}{Purple!60}
\DTLsetpiesegmentcolor{7}{LimeGreen!60}
\DTLsetpiesegmentcolor{8}{Blue!60}
\DTLsetpiesegmentcolor{9}{Cyan!90}
\DTLsetpiesegmentcolor{10}{ForestGreen!60}

\begin{document}

\bstctlcite{IEEEexample:BSTcontrol}

\title{Automatically Tailoring Static Analysis\\to Custom Usage Scenarios}

\author{
	\IEEEauthorblockN{Muhammad Numair Mansur\IEEEauthorrefmark{1}, 
	Benjamin Mariano\IEEEauthorrefmark{2}, 
	Maria Christakis\IEEEauthorrefmark{1}, 
	Jorge A. Navas\IEEEauthorrefmark{3} and 
	Valentin W{\"u}stholz\IEEEauthorrefmark{4}}
	
	\IEEEauthorblockA{
		\IEEEauthorrefmark{1}MPI-SWS, Germany\\
		\texttt{\{numair,maria\}@mpi-sws.org}}
	\IEEEauthorblockA{
		\IEEEauthorrefmark{2}The University of Texas at Austin, USA\\
		\texttt{bmariano@cs.utexas.edu}}
	\IEEEauthorblockA{
		\IEEEauthorrefmark{3}SRI International, USA\\
		\texttt{jorge.navas@sri.com}}
	\IEEEauthorblockA{
		\IEEEauthorrefmark{4}ConsenSys, Germany\\
		\texttt{valentin.wustholz@consensys.net}}}

\maketitle

\begin{abstract}
In recent years, there has been significant progress in the
development and industrial adoption of static analyzers. Such
analyzers typically provide a large, if not huge, number of
configurable options controlling the precision and performance of the
analysis. A major hurdle in integrating static analyzers in the
software-development life cycle is tuning their options to custom
usage scenarios, such as a particular code base or certain resource
constraints.

In this paper, we propose a technique that automatically tailors a
static analyzer, specifically an abstract interpreter, to the code
under analysis and any given resource constraints. We implement this
technique in a framework called \tool, which we use to perform an
extensive evaluation on real-world benchmarks. Our experiments show
that the configurations generated by \tool are vastly better than the
default analysis options, vary significantly depending on the code
under analysis, and most remain tailored to several subsequent
code versions.
\end{abstract}

\section{Introduction}
\label{sec:Introduction}

\emph{Static analysis} inspects code, without running it, in order to
prove properties or detect bugs.  Typically, static analysis
approximates the behavior of the code, for instance, because checking
the correctness of most properties is undecidable. \emph{Performance}
is another important reason for this approximation. In general, the
closer the approximation is to the actual behavior of the code, the
less efficient and the more \emph{precise} the analysis is, that is,
the fewer false positives it reports. For less tight approximations,
the analysis often becomes more efficient but less precise.

Recent years have seen tremendous progress in both the development and
industrial adoption of static analyzers. Notable successes include
Facebook's Infer~\cite{CalcagnoDistefano2015,CalcagnoDistefano2011}
and AbsInt's Astr\'ee~\cite{BlanchetCousot2003}.
Many popular analyzers, such as these, are based on \emph{abstract
  interpretation}~\cite{CousotCousot1977}, a technique that abstracts
the concrete program semantics and reasons about its abstraction. In
particular, program states are abstracted as elements of
\emph{abstract domains}.
Most abstract interpreters offer a wide range of abstract domains that
impact the precision and performance of the analysis. For
instance, the Intervals domain~\cite{CousotCousot1976} is typically
faster but less precise than Polyhedra~\cite{CousotHalbwachs1978},
which captures linear inequalities among any number of variables.

In addition to the domains, abstract interpreters usually provide a
large number of other options, for instance, whether backward analysis
should be enabled or how quickly a fixpoint should be reached. In
fact, the sheer number of option combinations (over 6M in our
experiments) is bound to overwhelm users, especially non-expert
ones. To make matters worse, the best option combinations may vary
significantly depending on the code under analysis and the resources,
such as time or memory, that users are willing to spend.

In light of this, we suspect that most users resort to using the
default options that the analysis designer pre-selected for
them. However, these options are definitely not suitable for all
code. Moreover, they do not adjust to different stages of software
development, e.g., running the analysis in the editor should be much
faster than running it in a continuous integration (CI) pipeline,
which in turn should be much faster than running it prior to a major
release.
The alternative of enabling the (in theory) most precise analysis can
be even worse, since in practice it often runs out of time or memory
as we show in our experiments.
As a result, the widespread adoption of abstract interpreters is
severely hindered, which is unfortunate since they constitute an
important class of practical static analyzers.

\textbf{Our approach.}
To address this issue, we present the first technique that
automatically tailors a generic abstract interpreter to a custom usage
scenario. With the term \emph{custom usage scenario}, we refer to a
particular piece of code and specific resource constraints. The key
idea behind our technique is to phrase the problem of customizing the
abstract-interpretation configuration to a given usage scenario as an
optimization problem. Specifically, different configurations are
compared using a cost function that penalizes those that prove fewer
properties or require more resources. This cost function can guide the
configuration search of a wide range of existing optimization
algorithms.

We implement our technique in a framework called \tool,
which configures a given abstract interpreter for a given usage
scenario using a given optimization algorithm. As a result, \tool
enables the abstract interpreter to prove as many properties as
possible within the resource limit without requiring any domain
expertise on behalf of the user.

Using \tool, we find that tailored configurations vastly outperform
the default options pre-selected by the analysis designers. In fact,
we show that this is possible even with very simple optimization
algorithms. Our experiments also demonstrate that tailored
configurations vary significantly depending on the usage scenario---in
other words, there cannot be a single configuration that fits all
scenarios. Finally, most of the generated configurations remain tailored
to several subsequent code versions, suggesting that re-tuning is
only necessary after major code changes.

\textbf{Contributions.}
We make the following contributions:
\begin{enumerate}
\item We present the first technique for automatically tailoring
  abstract interpreters to custom usage scenarios.
\item We implement our technique in a framework called \tool.
\item Using a state-of-the-art abstract interpreter with millions of
  configurations, we show the effectiveness of \tool on real-world
  benchmarks.
\end{enumerate}

\textbf{Outline.}
In the next section, we give a high-level overview of our technique
and framework. \secref{sec:abstract-interpreters} provides background
on the generic architecture of abstract
interpreters. \secref{sec:Technique} describes our technique in
detail, and \secref{sec:Experiments} presents our experimental
evaluation. We discuss related work in \secref{sec:RelatedWork} and
conclude in \secref{sec:Conclusion}.

\section{Overview}
\label{sec:Overview}

We now illustrate the workflow and tool architecture of \tool and
provide examples of its effectiveness.

\textbf{Terminology.}
In the following, we refer to an abstract domain with all its options
(e.g., enabling backward analysis or more precise treatment of arrays
etc.) as an \emph{ingredient}.

As discussed earlier, abstract interpreters typically provide a large
number of such ingredients. To make matters worse, it is also possible
to combine different ingredients into a sequence (which we call a
\emph{recipe}) such that more properties are verified than with
individual ingredients. For example, a user could configure the
abstract interpreter to first use Intervals to verify as many
properties as possible and then use Polyhedra to attempt verification
of any remaining properties.
Of course, the number of possible configurations grows exponentially
in the length of the recipe (over 6M in our experiments for recipes up
to length 3).


\textbf{Workflow.} The high-level architecture of \tool is shown in
\figref{fig:overview}.
It takes as input the code to be analyzed (i.e., any program, file,
function, or fragment), a user-provided resource limit, and optionally
an optimization algorithm. We focus on time as the constrained
resource in this paper, but our technique could be easily extended to
other resources, such as memory.

The optimization engine relies on a recipe generator to generate a
fresh recipe. To assess its quality in terms of precision and
performance, the recipe evaluator computes a cost for the recipe. The
cost is computed by evaluating how precise and efficient the abstract
interpreter is for the given recipe. This cost is used by the
optimization engine to keep track of the best recipe so far, i.e., the
one that proves the most properties in the least amount of time. \tool
repeats this process for a given number of iterations to sample
multiple recipes and returns the recipe with the lowest cost.

Zooming in on the evaluator, a recipe is processed by invoking the
abstract interpreter for each ingredient. After each analysis (i.e.,
one ingredient), the evaluator collects the new verification results,
that is, the verified assertions. All verification results that have
been achieved so far are subsequently shared with the analyzer when it
is invoked for the next ingredient. Verification results are shared by
converting all verified assertions into assumptions. After processing
the entire recipe, the evaluator computes a cost for the recipe, which
depends on the number of unverified assertions and the total analysis
time.

In general, there might be more than one recipe tailored to a
particular usage scenario. Na\"ively, finding one requires searching
the space of all recipes. \secref{subsec:Algorithms} discusses several
optimization algorithms for performing this search, which \tool
already incorporates in the optimization engine.

\begin{figure}[t]
\centering
\scalebox{0.80}{
  \begin{tikzpicture}[align=center, node distance=1.5cm]

    \node[draw=none] (P) at (0,0) {code + resources +\\optimization algorithm};
    \node[draw, rounded corners=3, fill=white, draw=black!50, below of=P, xshift=3.5cm, yshift=-0.6cm, dashed] (AD) {
    \begin{minipage}[b][2.5cm]{10.2cm}
    \tool
    \end{minipage}};
    \node[draw, rounded corners=3, fill=NavyBlue!30, draw=black!50, below of=P, yshift=0cm, text width=1.8cm] (F) {Optimization\\Engine};
    \node[draw, rounded corners=3, fill=Crimson!40, draw=black!50, right of=F, xshift=2.0cm, text width=1.8cm] (X) {Recipe Generator};

    \node[draw, rounded corners=3, fill=Green!40, draw=black!50, right of=X, xshift=2.0cm, text width=1.8cm] (N) {Recipe Evaluator};

    \node[draw, rounded corners=3, fill=DarkOrange!40, draw=black!50, below of=N, yshift=-1.3cm, text width=1.8cm] (L) {Static\\Analyzer};
    \node[draw=none, below of=F, yshift=-1.1cm, text width=2.5cm] (T) {tailored recipe};
    \draw[thick, ->, shorten >=1pt] (P) -- (F);
    \draw[thick, ->, shorten >=1pt] (F) -- (T);
    \draw[thick, ->, shorten >=1pt] (F) -- (X); 
    \draw[thick, ->, shorten >=1pt] (X) -- (N) node[midway, anchor=center, fill=white] {recipe};
    \draw[thick, ->, shorten >=1pt, out=-155, in=-25, looseness=0.9] (N.south west) to node[midway, anchor=center, fill=white, yshift=-0.35cm, xshift=0.1cm] {cost} (F.south east);

    \begin{scope}[transform canvas={xshift=.8em}]
  		\draw[thick, ->, shorten >=1pt] (N) to node[midway, anchor=center, xshift=0.6cm] {ingr. +\\current\\results} (L);
	\end{scope}

	\begin{scope}[transform canvas={xshift=-.8em}]
  		\draw[thick, ->, shorten >=1pt] (L) to node[midway, anchor=center, xshift=-0.6cm] {new\\results} (N);
	\end{scope}

  \end{tikzpicture}
}
\vspace{-0em}
\caption{Overview of \tool.}
\label{fig:overview}
\vspace{-0em}
\end{figure}

\textbf{Examples.}
As an example, let us consider the usage scenario where a user runs
the \crab abstract interpreter~\cite{GurfinkelKahsai2015} in their
editor for instant feedback during code development. This means that
the allowed time limit for the analysis is very short, say, 1~sec. Now
assume that the code under analysis is a program
file\footnote{\scriptsize\url{https://github.com/FFmpeg/FFmpeg/blob/master/libavformat/idcin.c}}
of the multimedia processing tool \textsc{ffmpeg}, which is used to
evaluate the effectiveness of \tool in our experiments. In this file,
\crab checks 45 assertions for common bugs, namely, division by zero,
integer overflow, buffer overflow, and use after free.

Analysis of this file with the default \crab configuration takes
0.35~sec to complete. In this time, \crab proves 17 assertions and
emits 28 warnings about the properties that remain unverified. For
this usage scenario, \tool is able to tune the abstract-interpreter
configuration such that the analysis time is 0.57~sec and the number
of verified properties increases by 29\% (i.e., 22 assertions are
proved). Note that the tailored configuration uses a completely
different abstract domain than the one used in the default
configuration.  As a result, the verification results are
significantly better, but the analysis takes slightly longer to
complete (although remaining within the specified time limit). In
contrast, enabling the most precise analysis in \crab verifies 26
assertions but takes over 6~min to complete, which by far exceeds the
time limit imposed by the specified usage scenario.

While it takes \tool 4.5~sec to find the above configuration, this is
time well invested; the configuration can be re-used for several
subsequent code versions. In fact, in our experiments, we show that
generated configurations can remain tailored for at least up to 50
subsequent commits to a file under version control. Given that changes
in the editor are typically much more incremental, we expect that no
re-tuning would be necessary at all during an editor
session. Re-tuning may be beneficial after major changes to the code
under analysis and can happen offline, e.g., between editor
sessions, or in the worst case overnight.

As another example, consider the usage scenario where \crab is
integrated in a CI pipeline. In this scenario, users should be able to
spare more time for analysis, say, 5~min. Here, let us assume that the
analyzed code is a program
file\footnote{\scriptsize\url{https://github.com/curl/curl/blob/master/lib/cookie.c}}
of the \textsc{curl} tool for transferring data by URL, which is also
used in our evaluation. The default \crab configuration takes 0.23~sec
to run and only verifies 2 out of 33 checked assertions. \tool is able
to find a configuration that takes 7.6~sec and proves 8 assertions. In
contrast, the most precise configuration does not terminate even after
15 min.

Both usage scenarios demonstrate that, even when users have more time
to spare, the default configuration cannot take advantage of it to
improve the verification results. At the same time, the most precise
configuration is completely impractical since it does not respect the
resource constraints imposed by these scenarios.


\section{Background: A Generic Abstract Interpreter}
\label{sec:abstract-interpreters}

\begin{figure}[t]
\centering
\scalebox{0.80}{
  \begin{tikzpicture}[scale=.8, node distance=0.5cm, align=center,
      outer/.style={ 
                     },
      inner/.style={rounded corners=3,draw=black!50
        }
    ]

    \node[outer] (mem) {Memory Domain};
    \draw [draw=black!50] (mem.north west) rectangle +(8.95cm, -3cm);
    \node[draw,inner,fill=Blue!30,below of=mem, xshift=0.5cm] (ptr) {Pointer Domains};
    \node[draw,inner,fill=Blue!30,right=1cm of ptr] (array) {Array Domains};
    \node[draw,inner,thick,fill=Blue!30,xshift=1.2cm, below=0.5cm of ptr] (num) {Logico/Numerical Domains};

    \draw[thick, <->] (ptr) -- (array);
    \draw[thick, ->] (ptr) -- (num);
    \draw[thick, ->] (array) -- (num);

    \node [outer, below of=num, xshift=-2.6cm,yshift=-1cm] (engine) {Analysis Engine};
    \draw [draw=black!50] (engine.north west) rectangle +(7.4cm, -4.95cm);
    \draw [draw=black!50,dashed,rounded corners=3] ([xshift=-1.25cm,yshift=0.25cm]mem.north west) rectangle +(11.3cm, -9.25cm);

    \node[draw,inner,fill=DarkOrange!30,text width=4cm,xshift=1.8cm,below of=engine] (abstr) {Abstract Transformers};
    \node[draw,inner,fill=DarkOrange!30,text width=4cm,yshift=-0.01cm, below of=abstr] (fixpo) {Fixpoint Computation};
    \node[draw,inner,fill=DarkOrange!30,text width=4cm,yshift=-0.26cm, below of=fixpo] (direction) {Forward/Backward Analysis};
    \node[draw,inner,fill=DarkOrange!30,text width=4cm,yshift=-0.265cm, below of=direction] (inter) {Intra/Inter-Proc. Analysis};
    \node[draw,inner,fill=DarkOrange!30,text width=4cm,yshift=-0.055cm, below of=inter] (prog) {CFG/Call-Graph Builder};

    \coordinate (engine1) at ([shift={(0cm,0.785cm)}]engine.north);
    \draw[thick, <->] (engine.north) -- (engine1);

    \node[draw,outer,fill=Red!30,text width=2cm,text height=0.3cm,text depth=0.5cm,right of=direction,xshift=4.0cm] (checker) {Assertion Checker};
    \draw[thick, <->] (direction.east) ++(0.875,0) -- (checker) ;

    \coordinate (checker1) at ([shift={(0cm,2.7cm)}]checker.north);
    \draw[thick, <->] (checker.north) -- (checker1);
  \end{tikzpicture}}
  \vspace{0.0em}
  \caption{Generic architecture of an abstract interpreter.}
  \vspace{0.0em}
  \label{fig:interpreter}
\end{figure}
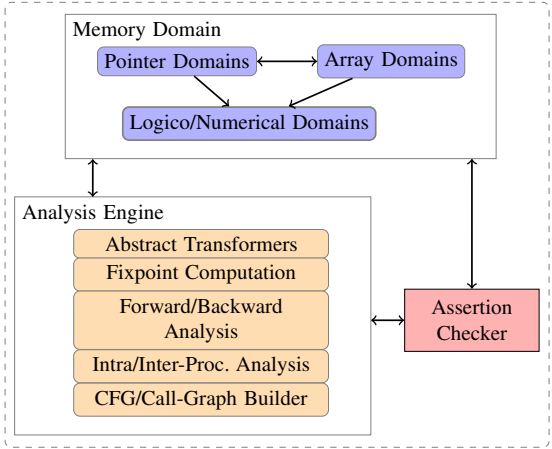

Many successful abstract interpreters (e.g.,
Astr\'ee~\cite{BlanchetCousot2003}, C Global
Surveyor~\cite{VenetBrat2004},
Clousot~\cite{FahndrichLogozzo2010}, \crab~\cite{GurfinkelKahsai2015},
IKOS~\cite{BratNavas2014}, Sparrow~\cite{OhHeo2012}, and
Infer~\cite{CalcagnoDistefano2015}) follow the generic architecture in
Fig.~\ref{fig:interpreter}.
In this section, we describe the main components of such a generic
abstract interpreter.

\textbf{Memory domain.}
Analysis of low-level languages such as C and LLVM-bitcode requires
reasoning about pointers. It is, therefore, common to design
a \emph{memory domain}~\cite{Mine2006-Fields} that can simultaneously
reason about pointer aliasing, memory contents, and numerical
relations between them.

\emph{Pointer domains} resolve aliasing
between pointers, and \emph{array domains} reason about memory
contents.
In particular, array domains can reason about individual memory locations (cells), infer
universal properties over multiple cells, or both. Typically, reasoning about individual
cells trades performance for precision unless there are very few array elements
(e.g.,~\cite{GershuniAmit2019,Mine2006-Fields}). In contrast, reasoning about multiple
memory locations (\emph{summarized cells}) trades precision for performance. In our
evaluation, we use \emph{Array smashing} domains~\cite{BlanchetCousot2003} that abstract
different array elements into a single summarized cell.

\emph{Logico-numerical domains} infer relationships between
program and \emph{synthetic} variables, introduced by the pointer and
array domains, e.g., summarized cells. Next, we introduce domains
typically used for proving the absence of runtime errors in low-level
languages.

\emph{Boolean domains} (e.g., flat Boolean, BDDApron~\cite{BDDApron})
reason about Boolean variables and expressions. \emph{Non-relational
domains} (e.g., Intervals~\cite{CousotCousot1976},
Congruence~\cite{Granger1989}) do not track relations among different
variables, in contrast to \emph{relational domains} (e.g.,
Equality~\cite{Karr1976}, Zones~\cite{Mine2002},
Octagons~\cite{Mine2006-Octagons},
Polyhedra~\cite{CousotHalbwachs1978}). Due to their increased
precision, relational domains are typically less efficient than
non-relational ones. \emph{Symbolic domains} (e.g., Congruence
closure~\cite{ChangLeino2005}, Symbolic
constant~\cite{Mine2006-Precision}, Term~\cite{GangeNavas2016-Domain})
abstract complex expressions (e.g., non-linear) and external library
calls by uninterpreted functions. \emph{Non-convex domains} express
disjunctive invariants. For instance, the DisInt
domain~\cite{FahndrichLogozzo2010} extends Intervals to a finite
disjunction; it retains the scalability of the Intervals domain by
keeping only non-overlapping intervals. On the other hand, the Boxes
domain~\cite{GurfinkelChaki2010} captures arbitrary Boolean
combinations of intervals, which can often be expensive.

\textbf{Fixpoint computation.}
To ensure termination of the fixpoint computation, Cousot and Cousot
introduce
\emph{widening}~\cite{CousotCousot1977,CousotCousot1992-Galois}, which usually incurs a loss of precision. There
are three common strategies to reduce this precision loss, which
however sacrifice efficiency.
First, \emph{delayed widening}~\cite{BlanchetCousot2003} performs a
number of initial fixpoint-computation iterations in the hope of
reaching a fixpoint before resorting to
widening. Second, \emph{widening with
thresholds}~\cite{LakhdarChaouchJeannet2011,MihailaSepp2013} limits
the number of program expressions (thresholds) that are used when
widening.  The third strategy consists in
applying \emph{narrowing}~\cite{CousotCousot1977,CousotCousot1992-Galois}
a certain number of times.

\textbf{Forward and backward analysis.}
Classically, abstract interpreters analyze code by propagating
abstract states in a \emph{forward} manner. However, abstract
interpreters can also perform \emph{backward} analysis to compute the
execution states that lead to an assertion violation.
Cousot and
Cousot~\cite{CousotCousot1992-Applications,CousotCousot1999} define
a \emph{forward-backward refinement} algorithm in which a forward
analysis is followed by a backward analysis until no more refinement
is possible. The backward analysis uses invariants computed by the
forward analysis, while the forward analysis does not explore states
that cannot reach an assertion violation based on the backward
analysis. This refinement is more precise than forward analysis alone,
but it may also become very expensive.

\textbf{Intra- and inter-procedural analysis.}
An \emph{intra-procedural} analysis analyzes a function ignoring the
information (i.e., call stack) that flows into that function, while
an \emph{inter-procedural} analysis considers all the flows among
functions. The former is much more efficient and easy to parallelize,
but the latter is usually more precise.

\section{Our Technique}
\label{sec:Technique}

In this section, we describe the main components of \tool in detail;
\secsref{subsec:Tunning}, \ref{subsec:CostFunction},
\ref{subsec:Algorithms} explain the optimization engine, recipe
evaluator, and recipe generator from \figref{fig:overview}.

\subsection{Recipe Optimization}
\label{subsec:Tunning}

\algoref{alg:optimize-algorithm} implements the optimization
engine. In addition to the code $\prog$ and the resource limit
$\rmax$ it also takes
as input the maximum length of the generated recipes
$\lmax$ (i.e., the maximum number of ingredients), a
function to generate new recipes \textsc{GenerateRec} (i.e., the recipe
generator from \figref{fig:overview}), and four other parameters,
which we explain later.

A tailored recipe is found in two phases. The first phase aims to find
the best abstract domain for each ingredient, while the second tunes
the remaining analysis settings for each ingredient (e.g., whether
backward analysis should be enabled). Parameters
$\idom$ and $\iset$ control the number
of iterations of each phase. Note that we start with a search for the
best domains since they have the largest impact on the precision and
performance of the analysis.

During the first phase, the algorithm initializes the best recipe
$\best{rec}$ with an initial recipe $\init{rec}$ (line~3). The cost of this
recipe is evaluated with function \textsc{Eval}, which implements
the recipe-evaluator component from \figref{fig:overview}.
The subsequent nested loop (line~5) samples a number of recipes,
starting with the shortest recipes ($l$ := $1$) and ending with the
longest recipes ($l$ := $\lmax$). The inner loop generates
$\idom$ ingredients for each ingredient in the recipe (i.e., $\idom \cdot l$
total iterations) by invoking
function \textsc{GenerateRec}, and in case a recipe with lower cost is
found, it updates the best recipe (lines~9--10).
Several optimization algorithms, such as hill climbing and simulated
annealing, search for an optimal result by mutating some of the
intermediate results. Variable $\cur{rec}$ stores intermediate
recipes to be mutated, and function \textsc{Accept} decides when to
update it (lines~11--12).

As explained earlier, the purpose of the first phase is to identify
the best sequence of abstract domains. The second phase (lines~13--18)
focuses on tuning the other settings of the best recipe so
far. This is done by randomly mutating the best recipe via
\textsc{MutateSettings} (line~15), and updating
the best recipe if better settings are found (lines~17--18).
After exploring $\iset$ random settings, the best recipe is
returned to the user (line~19).

\IncMargin{0.1cm}
\begin{algorithm}[t!]
\small
\SetKwProg{Fn}{Function}{is}{end}
\Fn{\normalfont{\textsc{Optimize}($\prog$, $\rmax$, $\lmax$, $\idom$, $\iset$, $\init{rec}$, \textsc{GenerateRec}, \textsc{Accept})} \ }{
  // \textit{Phase 1 (optimize domains)}\\
  $\best{rec}$ := $\cur{rec}$ := $\init{rec}$\\
  $\best{\cost}$ := $\cur{\cost}$ := \textsc{Eval}($\prog$, $\rmax$, $\best{rec}$)\\
  \For{$l$ \normalfont{:=} $1$ \normalfont{\textbf{to}} $\lmax$}{
    \For{$i$ \normalfont{:=} $1$ \normalfont{\textbf{to}} $\idom \cdot l$}{
      $\nex{rec}$ := \textsc{GenerateRec}($\cur{rec}$, $l$)\\
      $\nex{\cost}$ := \textsc{Eval}($\prog$, $\rmax$, $\nex{rec}$)\\
      \If{$\nex{\cost} < \best{\cost}$}{
        $\best{rec}, \best{\cost}$ := $\nex{rec}$, $\nex{\cost}$
      }
      \If{\textsc{Accept}\normalfont{(}$\cur{\cost}$, $\nex{\cost}$\normalfont{)}}{
        $\cur{rec}, \cur{\cost}$ := $\nex{rec}$, $\nex{\cost}$
      }
    }
  }
  // \textit{Phase 2 (optimize settings)}\\
  \For{$i$ \normalfont{:=} $1$ \normalfont{\textbf{to}} $\iset$}{
    $\mut{rec}$ := \textsc{MutateSettings}($\best{rec}$)\\
    $\mut{\cost}$ := \textsc{Eval}($\prog$, $\rmax$, $\mut{rec}$)\\
    \If{$\mut{\cost} < \best{\cost}$}{
      $\best{rec}, \best{\cost}$ := $\mut{rec}$, $\mut{\cost}$
    }
  }
  \Return{$\best{rec}$}
}
\caption{\textbf{Optimization engine.}}
\label{alg:optimize-algorithm}
\end{algorithm}

\subsection{Recipe Evaluation}
\label{subsec:CostFunction}

The recipe evaluator from \figref{fig:overview} uses a cost function
to determine the quality of a fresh recipe with respect to the
precision and performance of the abstract interpreter. This design is
motivated by the fact that analysis imprecision and inefficiency are
among the top pain points for users~\cite{ChristakisBird2016}.

Therefore, the cost function depends on the number of generated
warnings $w$ (that is, the number of unverified assertions), the total
number of assertions in the code $w_{\mathit{total}}$, the resource
consumption $r$ of the analyzer, and the resource limit
$\rmax$ imposed on the analyzer:
$$
cost(w, w_{\mathit{total}}, r, \rmax) =
\begin{cases}
\dfrac{w + \dfrac{r}{\rmax}}{w_{\mathit{total}}},  & \text{if }r \leq \rmax \\
\infty, & \text{otherwise}
\end{cases}
$$

Note that $w$ and $r$ are measured by invoking the abstract
interpreter with the recipe under evaluation.
The cost function evaluates to a lower cost for recipes that improve
the precision of the abstract interpreter (due to the term $w /
w_{\mathit{total}}$). In case of ties, the term $r / \rmax$
causes the function to evaluate to a lower cost for recipes that
result in a more efficient analysis.
In other words, for two recipes resulting in equal precision, the one
with the smaller resource consumption is assigned a lower cost.
When a recipe causes the analyzer to exceed the resource limit, it is
assigned infinite cost.

\subsection{Recipe Generation}
\label{subsec:Algorithms}

In the literature, there is a broad range of optimization algorithms
for different application domains. To demonstrate the generality and
effectiveness of \tool, we instantiate it with four adaptations of
three well-known optimization algorithms, namely random
sampling~\cite{Matyas1965}, hill climbing (with regular
restarts)~\cite{ArtificialIntelligence}, and simulated
annealing~\cite{MetropolisRosenbluth1953,KirkpatrickGelatt1983}. Here,
we describe these algorithms in detail, and in
\secref{sec:Experiments}, we evaluate their effectiveness.

Before diving into the details, let us discuss the suitability of
different kinds of optimization algorithms for our domain. There are
algorithms that leverage mathematical properties of the function to be
optimized, e.g., by computing derivatives as in Newton's iterative
method. Our cost function, however, is evaluated by running an
abstract interpreter, and thus, it is not differentiable or
continuous. This constraint makes such analytical algorithms
unsuitable.  Moreover, evaluating our cost function is expensive,
especially for precise abstract domains such as Polyhedra. This makes
algorithms that require a large number of samples, such as genetic
algorithms, less practical.

Now recall that \algoref{alg:optimize-algorithm} is parametric in how
new recipes are generated (\textsc{GenerateRec}) and accepted for further
mutations (\textsc{Accept}). Instantiations of these functions
essentially constitute our search strategy for a tailored recipe. In
the following, we discuss four such instantiations.
Note that, in theory, the order of recipe ingredients matters. This is
because any properties verified by one ingredient are converted into
assumptions for the next, and different assumptions may lead to
different verification results. Therefore, all our instantiations are
able to explore different ingredient orderings.

\textbf{Random sampling.}
Random sampling (\textsc{rs}) just generates random recipes of a
certain length. Function \textsc{Accept} always returns
$\mathit{false}$ as each recipe is generated from scratch, and not as
a result of any mutations.

\textbf{Domain-aware random sampling.}
\textsc{rs} might generate recipes containing two or more abstract
domains of comparable precision. For instance, the Octagons domain is
typically strictly more precise than Intervals. As a result, a recipe
consisting of these domains is essentially equivalent to a recipe
containing only Octagons.

Now, assume that we have a partially ordered set (poset) of domains
that defines their ordering in terms of precision. An example of such
a poset for a particular abstract interpreter is shown in
\figref{fig:lattice}. An optimization algorithm can then leverage this
information to reduce the search space of possible recipes.
Given such a poset, we therefore define domain-aware random sampling
(\textsc{dars}), which randomly samples recipes that do not contain
abstract domains of comparable precision. Again, \textsc{Accept}
always returns $\mathit{false}$.

\textbf{Simulated annealing.}
Simulated annealing (\textsc{sa}) searches for the best recipe by
mutating the current recipe $\cur{rec}$ in
\algoref{alg:optimize-algorithm}.  The resulting recipe
($\nex{rec}$), if accepted on line~12, becomes the new recipe
to be mutated. \algoref{alg:recipe-generation-algorithm} shows
an instantiation of \textsc{GenerateRec},
which mutates a given recipe such
that the poset precision constraints are satisfied (i.e., there are no
domains of comparable precision). A recipe is mutated either by adding
new ingredients with 20\% probability or by modifying existing
ones with 80\% probability (line~2).
The probability of adding ingredients is lower to keep recipes short.

When adding a new ingredient (lines~4--5),
\algoref{alg:recipe-generation-algorithm}
calls \textsc{RandPosetLeastInc}, which considers all domains that
are incomparable with the domains in the recipe. Given this set, it
randomly selects from the domains with the least precision to avoid
adding overly expensive domains.
When modifying a random ingredient in the recipe (lines~7--16),
the algorithm can replace its domain with one of three
possibilities: a domain that is immediately more precise (i.e., not transitively)
in the poset (via \textsc{PosetGT}), a domain that is immediately
less precise (via \textsc{PosetLT}), or an incomparable domain with
the least precision (via \textsc{RandPosetLeastInc}).
In case the resulting recipe does not satisfy the
poset precision constraints, our algorithm retries to mutate the
original recipe (lines~17--18).

For simulated annealing, function $\textsc{Accept}$ returns
$\mathit{true}$ if the new cost (for the mutated recipe) is less than
the current cost. It also accepts recipes whose cost is higher with a
certain probability, which is inversely proportional to the cost
increase as well as the number of recipes explored so far. In other
words, recipes with a small cost increase are likely to be accepted,
especially toward the beginning of the exploration.

\textbf{Hill climbing.}
Our instantiation of hill climbing (\textsc{hc}) performs regular
restarts. In particular, it starts with a randomly generated recipe
that satisfies the poset precision constraints, generates 10 new valid
recipes, and restarts
with a random recipe.
\textsc{Accept} returns $\mathit{true}$ only if the new cost
is lower than the best cost, which is equivalent to the current
cost.

\begin{algorithm}[t!]
\small
\DontPrintSemicolon
\SetKwProg{Fn}{Function}{is}{end}

\Fn{\normalfont{\textsc{GenerateRec}($\rec$, $\lmax$) } \ }{
  $\act$ := \textsc{RandAct}(\{$\add$: 0.2, $\modfy$: 0.8\}))\\
  \eIf{$\act = \add \land \textsc{Len}(\rec) < \lmax$}{
    $\new{ingr}$ := \textsc{RandPosetLeastInc}($\rec$)\\
    $\mut{rec}$ := \textsc{AddIngr}($\rec$, $\new{ingr}$)\\
  }{
    $\ingr$ := \textsc{RandIngr}($\rec$)\\
    $\modact$ := \textsc{RandAct}(\{$\pgt$: 0.5, $\plt$: 0.3, $\inc$: 0.2\})\\
    \uIf{$\modact = \pgt$}{
      $\new{ingr}$ := \textsc{PosetGT}($\ingr$)
    }
    \uElseIf{$\modact = \plt$}{
      $\new{ingr}$ := \textsc{PosetLT}($\ingr$)      
    }
    \Else{
      $\mathrm{rec}_{\mathrm{rem}}$ := \textsc{RemoveIngr}($\rec$, $\ingr$)\\
      $\new{ingr}$ := \textsc{RandPosetLeastInc}($\mathrm{rec}_{\mathrm{rem}}$)      
    }
    $\mut{rec}$ := \textsc{ReplaceIngr}($\rec$, $\ingr$, $\new{ingr}$)\\
  }
  \If{\normalfont{$\neg$\textsc{PosetCompat}($\mut{rec}$)}}{
    $\mut{rec}$ := \textsc{GenerateRec}($\rec$, $\lmax$)
  }

  \Return{$\mut{rec}$}

}

\caption{\textbf{A recipe-generator instantiation.}}
\label{alg:recipe-generation-algorithm}
\end{algorithm}

\section{Experimental Evaluation}
\label{sec:Experiments}

To evaluate our technique, we aim to answer the following research
questions:

\begin{description}
\item[RQ1:] Is our technique effective in finding tailored recipes for
  different usage scenarios?
\item[RQ2:] Are the tailored recipes optimal?
\item[RQ3:] How diverse are the tailored recipes?
\item[RQ4:] How resilient are the tailored recipes to code changes?
\end{description}

\subsection{Implementation}
\label{subsec:crab}


We implemented \tool by extending \crab~\cite{GurfinkelKahsai2015}, a parametric
framework for modular construction of abstract
interpreters\footnote{\crab is available at
  \scriptsize\url{https://github.com/seahorn/crab}.}.
We extended \crab with the ability to pass verification results
between recipe ingredients as well as with the four optimization
algorithms discussed in \secref{subsec:Algorithms}.

Tab.~\ref{tab:allowed-parameter-values} shows all settings and values
used in our evaluation. The first three settings refer to the
strategies discussed in \secref{sec:abstract-interpreters} for
mitigating the precision loss incurred by widening.
For the initial recipe, \tool uses Intervals
and the \crab default values for all other settings (in bold in the
table). To make the search more efficient, we selected a subset of all
possible setting values for our experiments. However, to ensure a
representative subset, we consulted with the \crab designer.

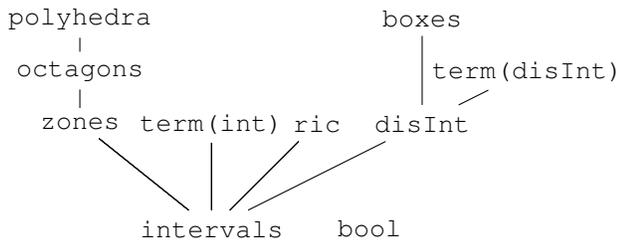
\begin{figure}[b]
\centering
\scalebox{1}{
\begin{tikzpicture}[scale=.7]

  \node (boxes) at (3,4) {\texttt{boxes}};
  \node (term-dis-int) at (5,3) {\texttt{term(disInt)}};
  \node (dis-int) at (3,2) {\texttt{disInt}};
  \node (poly) at (-3.5,4) {\texttt{polyhedra}};
  \node (oct) at (-3.5,3) {\texttt{octagons}};
  \node (zones) at (-3.5,2) {\texttt{zones}};
  \node (term-int) at (-1,2) {\texttt{term(int)}};
  \node (ric) at (1,2) {\texttt{ric}};
  \node (intervals) at (-1,0) {\texttt{intervals}};
  \node (bool) at (2,0) {\texttt{bool}};

  \draw (intervals) -- (ric) -- (intervals) -- (term-int) -- (intervals) -- (zones) -- (intervals) -- (dis-int) (zones) -- (oct) -- (poly) (dis-int) -- (term-dis-int) (dis-int) -- (boxes);

\end{tikzpicture}}

\caption{Comparing logico-numerical domains in \crab. A
  domain $d_1$ is less precise than $d_2$ if there is a path from
  $d_1$ to $d_2$ going upward, otherwise $d_1$ and $d_2$ are
  incomparable.}
\label{fig:lattice}
\end{figure}

\crab uses a DSA-based~\cite{GurfinkelNavas2017} pointer analysis and
can, optionally, reason about array contents using array smashing.
It offers a wide range of logico-numerical domains, shown in
Fig.~\ref{fig:lattice}. The \texttt{bool} domain is the flat Boolean
domain, \texttt{ric} is a reduced product of Intervals and Congruence,
and \texttt{term(int)} and \texttt{term(disInt)} are instantiations of
the Term domain with \texttt{intervals} and \texttt{disInt},
respectively.
%
%
Even though \crab provides a bottom-up inter-procedural analysis, our
evaluation uses the default intra-procedural analysis; in fact, most
static analyses deployed in real usage scenarios are intra-procedural
due to time constraints~\cite{ChristakisBird2016}.


\begin{table}[t]
\centering
\scalebox{1}{
\begin{tabular}{l|l}
\textbf{Setting} & \textbf{Possible Values}\\
\hline
\textsc{\delays}    & $\lbrace \textbf{1}, 2, 4, 8, 16 \rbrace$ \\
\textsc{\narrowings} & $\lbrace 1, \textbf{2}, 3, 4 \rbrace$ \\
\textsc{\thresholds}  & $\lbrace \textbf{0}, 10, 20, 30, 40 \rbrace$ \\
\textsc{Backward Analysis}     & $\lbrace \textbf{OFF}, \mathit{ON} \rbrace$ \\
\textsc{Array Smashing} & $\lbrace \mathit{OFF}, \textbf{ON} \rbrace$ \\
\textsc{Abstract Domains}   & all domains in Fig.~\ref{fig:lattice} \\
\end{tabular}
}
\caption{\crab settings and their possible values as used in our
  experiments. Default settings are shown in bold.}
\vspace{-1.5em}
\label{tab:allowed-parameter-values}
\end{table}

\subsection{Benchmark Selection}
\label{subsec:Benchmarks}

For our evaluation, we systematically selected popular
and (at some point) active C projects on GitHub. In particular, we
chose the six most starred C repositories with over 300 commits that
we could successfully build with the Clang-5.0 compiler. We give a
short description of each project in \tabref{tab:projects}.

For analyzing these projects using abstract interpretation, we needed
to introduce properties to be verified. For our purposes, we
instrumented these projects with four types of assertions that check
for common bugs; namely, division by zero, integer overflow, buffer
overflow, and use after free. Introducing assertions to check for
runtime errors such as these is common practice in program analysis
and verification.

As projects consist of different numbers of files, to avoid skewing
the results in favor of a particular project, we randomly and
uniformly sampled 20 LLVM-bitcode files from each project, for a total
of 120.
To ensure that each file was neither too trivial nor too difficult for
the abstract interpreter, we used the number of assertions as a
complexity indicator and only sampled files with at least 20
assertions and at most 100. Additionally, to guarantee all four
assertion types (listed above) were included and avoid skewing the
results in favor of a particular assertion type, we required that the
sum of assertions for each type was at least 70 across all files---this
exact number was largely determined by the benchmarks.

Overall, our benchmark suite of 120 files totals 1346 functions, 5557
assertions (on average 4 assertions per function), and 667927 LLVM
instructions (see \tabref{tab:benchmarks}).

\begin{table}[t]
\centering
\scalebox{1}{
\begin{tabular}{l|l}
\textbf{Project} & \textbf{Description}\\
\hline
\textsc{curl}    & Tool for transferring data by URL\\
\textsc{darknet} & Convolutional neural-network framework\\
\textsc{ffmpeg}  & Multimedia processing tool\\
\textsc{git}     & Distributed version-control tool\\
\textsc{php-src} & PHP interpreter\\
\textsc{redis}   & Persistent in-memory database\\
\end{tabular}
}
\caption{Overview of projects.}
\vspace{-1.5em}
\label{tab:projects}
\end{table}

\begin{table}[b]
\vspace{0em}
\centering
\scalebox{1}{
\begin{tabular}{l|S[table-format=2.0]|S[table-format=3.0]|S[table-format=3.0]}
\textbf{Project} & \textbf{Functions} & \textbf{Assertions}& \textbf{LLVM Instructions}\\
\hline
\textsc{curl}    & 306 & 787 & 50541 \\
\textsc{darknet} & 130 & 958 & 55847 \\
\textsc{ffmpeg}  & 103 & 888 & 27653 \\
\textsc{git}     & 218 & 768 & 102304 \\
\textsc{php-src} & 268 & 1031 & 305943 \\
\textsc{redis}   & 321 & 1125 & 125639 \\
\hline
\textbf{Total}   & 1346 & 5557 & 667927
\end{tabular}
}
\caption{Benchmark characteristics (20 files per project). The last
  three columns show the number of functions, assertions, and LLVM
  instructions in the analyzed files.}
\label{tab:benchmarks}
\end{table}




\subsection{Results}
\label{subsec:Results}

We now present our experimental results for each research question.
We performed all experiments on a 32-core \mbox{Intel}
\textregistered~Xeon \textregistered~E5-2667 v2 CPU~@~3.30GHz machine
with 264GB of memory, running Ubuntu 16.04.1 LTS.

\textbf{RQ1: Is our technique effective in finding tailored recipes for
  different usage scenarios?}
We instantiated \tool with the four optimization algorithms described
in \secref{subsec:Algorithms}: \textsc{rs}, \textsc{dars},
\textsc{sa}, and \textsc{hc}. We constrained the analysis time to
simulate two usage scenarios: 1~sec for instant feedback in the
editor, and 5~min for feedback in a CI pipeline. We compare \tool with
the default recipe (\textsc{def}), i.e., the default settings in \crab
as defined by its designer after careful tuning on a large set of
benchmarks over the years. \textsc{def} uses a combination of two
domains, namely, the reduced product of Boolean and Zones. The other
default settings are in Tab.~\ref{tab:allowed-parameter-values}.

For this experiment, we ran \tool with each optimization algorithm on
the 120 benchmark files, enabling optimization at the granularity of
files. Each algorithm was seeded with the same random seed. In
\algoref{alg:optimize-algorithm}, we restrict recipes to contain at
most $3$ domains ($\lmax = 3$) and set the number of
iterations for each phase to be $5$ and $10$ ($\idom =
5$ and $\iset = 10$).

%
\begin{figure}[t]
\centering
\scalebox{0.75}{
  \begin{tikzpicture}
    \begin{axis}[
      ybar,
      xticklabels={\textsc{dars},\textsc{rs},\textsc{hc},\textsc{sa},\textsc{def}},
      xtick=data,
      ylabel={\textbf{Number of verified assertions}},
      legend cell align=left,
      axis y line*=none,
      axis x line*=bottom,
      width=10cm,
      height=5.5cm,
      ymin=0,
      ymax=1400,
      ytick distance=700,
      bar width=0.6cm,
      area legend,
      nodes near coords,
      legend style={legend pos=north east,font=\small, draw=none, legend columns=-1}
      ]

      \pgfplotstableread{data/ver-assert-compariosn.dat}\unsoundness

      \addplot+[ybar, Blue, pattern color=Blue, pattern=north east lines]
          table[x expr=\coordindex, y=x5] from \unsoundness;

      \addplot+[ybar, Green, pattern color=Green, pattern=north west lines]
          table[x expr=\coordindex, y=x4] from \unsoundness;


      \legend{1sec, 5min} 
    \end{axis}
  \end{tikzpicture}
}
\vspace{-0.5em}
\caption{Comparison of the number of assertions verified with the best
  recipe generated by each optimization algorithm and with the default
  recipe, for varying timeouts.}
\label{fig:alg-comparison}
\vspace{-0.5em}
\end{figure}
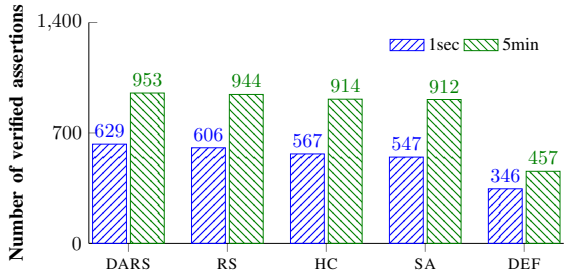

The results are presented in \figref{fig:alg-comparison}, which shows
the number of assertions that are verified with the best recipe found
by each algorithm as well as by the default recipe.  All algorithms
outperform the default recipe for both usage scenarios, verifying
almost twice as many assertions on average.

\begin{figure}[b]
\centering
\scalebox{0.75}{
\begin{tikzpicture}
\begin{axis}[
   xbar stacked,
	 ytick=data,
	 yticklabels={\textsc{dars}, \textsc{rs}, \textsc{hc}, \textsc{sa}},
     ylabel={\textbf{Optimization algorithms}},
     legend cell align=left,
     axis y line*=none,
     axis x line*=bottom,
     width=10cm,
     height=6cm,
     xmin=0,
     xmax=1100,
     ymax=5,
     xtick distance=500,
     bar width=0.6cm,
     area legend,
     legend style={legend pos=north east,font=\small, draw=none, legend columns=-1}
   ]

\addplot+[xbar, Green!70] plot coordinates {(575, 1) (578, 2) (535, 3) (526, 4)};

\addplot+[xbar, DarkOrange!70] plot coordinates {(378, 1) (366, 2) (379, 3) (386, 4)};

\addplot+[xbar, Red!70] plot coordinates {(79, 1) (91, 2) (78, 3) (71, 4)};

\legend{\tool, Both, \textsc{def}}

\coordinate (A) at (300,1);
\coordinate (B) at (300,2);
\coordinate (C) at (300,3);
\coordinate (D) at (300,4);

\coordinate (AA) at (750,1);
\coordinate (BB) at (750,2);
\coordinate (CC) at (750,3);
\coordinate (DD) at (750,4);

\coordinate (AAA) at (995,1);
\coordinate (BBB) at (990,2);
\coordinate (CCC) at (955,3);
\coordinate (DDD) at (950,4);

\end{axis}

\node at (A) {575};
\node at (B) {578};
\node at (C) {535};
\node at (D) {526};

\node at (AA) {378};
\node at (BB) {366};
\node at (CC) {379};
\node at (DD) {386};

\node at (AAA) {79};
\node at (BBB) {91};
\node at (CCC) {78};
\node at (DDD) {71};

\end{tikzpicture}
}
\vspace{-0.5em}
\caption{Comparison of the number of assertions verified by a tailored
  vs. the default recipe.}
\label{fig:compare-assertions}
\end{figure}
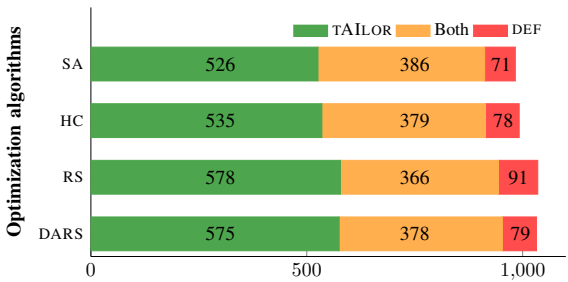

\figref{fig:compare-assertions} gives a more detailed comparison with
the default recipe for the time limit of 5~min. In particular, each
horizontal bar shows the total number of assertions verified by each
algorithm. The orange portion represents the assertions verified by
both the default recipe and the optimization algorithm, while the
green and red portions represent the assertions only verified by the
algorithm and default recipe, respectively. These results show that,
in addition to verifying hundreds of new assertions, \tool is able to
verify the vast majority of assertions proved by the default recipe,
regardless of optimization algorithm.

\begin{figure}[t]
\centering
\scalebox{0.75}{
  \begin{tikzpicture}
    \begin{axis}[
      ybar,
      xticklabels={\textsc{dars},\textsc{rs},\textsc{hc},\textsc{sa}},
      xtick=data,
      ylabel={\textbf{Total time (seconds)}},
      legend cell align=left,
      axis y line*=none,
      axis x line*=bottom,
      width=10cm,
      height=5.5cm,
      ymin=0,
      ymax=1600,
      ytick distance=800,
      bar width=0.6cm,
      area legend,
      nodes near coords,
      legend style={legend pos=north east,font=\small, draw=none, legend columns=-1}
      ]

      \pgfplotstableread{data/alg-total-time.dat}\unsoundness

      \addplot+[ybar, Blue, pattern color=Blue, pattern=north east lines]
          table[x expr=\coordindex, y=x5] from \unsoundness;

      \addplot+[ybar, Green, pattern color=Green, pattern=north west lines]
          table[x expr=\coordindex, y=x4] from \unsoundness;


      \legend{1sec, 5min} 
    \end{axis}
  \end{tikzpicture}
}
\vspace{-0.5em}
\caption{Comparison of the total time (in sec) that each algorithm
  requires for all iterations, for varying timeouts.}
\vspace{-0.5em}
\label{fig:alg-total-time}
\end{figure}
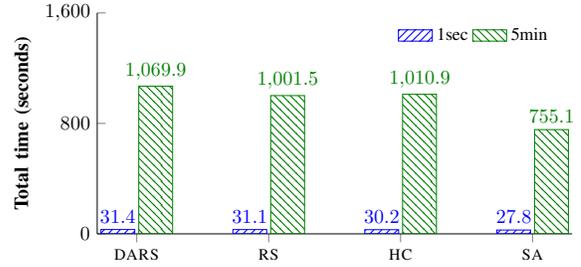

In \figref{fig:alg-total-time}, we show the total time that each
algorithm takes for all iterations. \textsc{dars} takes longer than
all others. This is due to generating more precise recipes thanks to
its domain knowledge. Such recipes typically take longer to run but
verify more assertions (as in \figref{fig:alg-comparison}).
On average, for all algorithms, \tool requires only $30$~sec to
complete all iterations for the $1$-sec timeout and $16$~min for the
$5$-min timeout.
As discussed in \secref{sec:Overview}, this tuning time can be spent
offline.

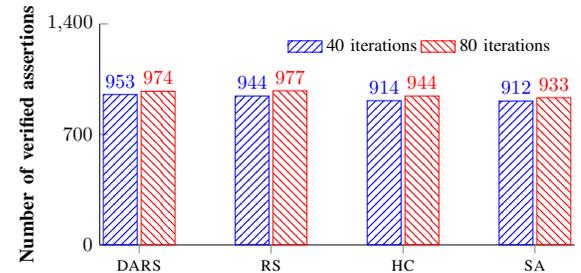
\begin{figure}[b]
\centering
\scalebox{0.75}{
  \begin{tikzpicture}
    \begin{axis}[
      ybar,
      xticklabels={\textsc{dars},\textsc{rs},\textsc{hc},\textsc{sa}},
      xtick=data,
      ylabel={\textbf{Number of verified assertions}},
      legend cell align=left,
      axis y line*=none,
      axis x line*=bottom,
      width=10cm,
      height=5.5cm,
      ymin=0,
      ymax=1400,
      ytick distance=700,
      bar width=0.6cm,
      area legend,
      nodes near coords,
      legend style={legend pos=north east,font=\small, draw=none, legend columns=-1}
      ]

      \pgfplotstableread{data/ver-assert-iter-compariosn.dat}\unsoundness

      \addplot+[ybar, Blue, pattern color=Blue, pattern=north east lines]
          table[x expr=\coordindex, y=x5] from \unsoundness;

      \addplot+[ybar, Red, pattern color=Red, pattern=north west lines]
          table[x expr=\coordindex, y=x4] from \unsoundness;

      \legend{40 iterations, 80 iterations}
    \end{axis}
  \end{tikzpicture}
}
\vspace{-0.5em}
\caption{Comparison of the number of assertions verified with
  the best recipe generated by the different optimization algorithms,
  for different numbers of iterations.}
\label{fig:alg-iter-comparison}
\end{figure}

\begin{figure}[t]
\centering
\scalebox{0.75}{
  \begin{tikzpicture}
    \begin{axis}[
      ybar,
      xticklabels={\textsc{dars},\textsc{rs},\textsc{hc},\textsc{sa}},
      xtick=data,
      ylabel={\textbf{Average number of iterations}},
      legend cell align=left,
      axis y line*=none,
      axis x line*=bottom,
      width=10cm,
      height=5.5cm,
      ymin=0,
      ymax=60,
      ytick distance=30,
      bar width=0.6cm,
      area legend,
      nodes near coords,
      legend style={legend pos=north east,font=\small, draw=none, legend columns=-1}
      ]

      \pgfplotstableread{data/ver-assert-iter-4080compariosn.dat}\unsoundness

      \addplot+[ybar, Blue, pattern color=Blue, pattern=north east lines]
          table[x expr=\coordindex, y=x5] from \unsoundness;

      \addplot+[ybar, Red, pattern color=Red, pattern=north west lines]
          table[x expr=\coordindex, y=x4] from \unsoundness;

      \legend{40 iterations, 80 iterations}
    \end{axis}
  \end{tikzpicture}
}
\vspace{-0.5em}
\caption{Comparison of the average number of iterations that each
  algorithm needs to find the best recipe, for varying numbers of
  iterations.}
\label{fig:alg-iter-best-iterations}
\vspace{-0.5em}
\end{figure}
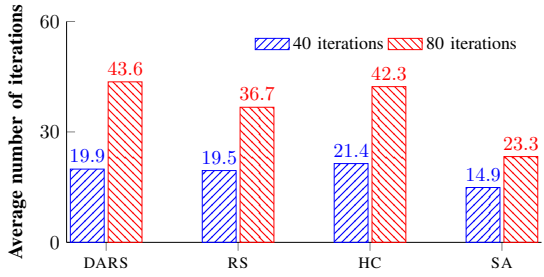

\figref{fig:alg-iter-comparison} compares the total number of
assertions verified by each algorithm when \tool runs for $40$
($\idom = 5$ and $\iset = 10$) and $80$
($\idom = 10$ and $\iset = 20$)
iterations.  The results show that only a relatively small number of
additional assertions are verified with 80 iterations. In fact, we
expect the algorithms to eventually converge on the number of verified
assertions, given the time limit and precision of the available
domains. \figref{fig:alg-iter-best-iterations} shows the number of
iterations required for each algorithm to find the best recipe, which
indicates that few assertions require more specialized recipes to be
proved and thus more iterations.

As \textsc{dars} performs best in this comparison, for the remaining
experiments, we only enable this algorithm with the $5$-min timeout
for simplicity. \\

\fbox{
  \begin{minipage}{0.87\linewidth}
    {\bf RQ1 takeaway:} \tool verifies nearly twice the assertions of
    the default recipe, regardless of optimization algorithm, timeout,
    or number of iterations. In fact, even very simple algorithms
    (such as \textsc{rs}) significantly outperform the default recipe.
  \end{minipage}
} \\


\textbf{RQ2: Are the tailored recipes optimal?}
To check the optimality of the tailored recipes, we compared them with
the most precise (and least efficient) \crab configuration.  It uses
the most precise domains from \figref{fig:lattice} (i.e.,
\texttt{bool}, \texttt{polyhedra}, \texttt{term(int)}, \texttt{ric},
\texttt{boxes}, and \texttt{term(disInt)}) in a recipe of 6
ingredients and assigns the most precise values to all other settings
from \tabref{tab:allowed-parameter-values}. We gave a 30-min timeout
to this recipe.

For 21 out of 120 files, the most precise recipe ran out of memory
(264GB). For 86 files, it terminated within 5~min, and for 13, it took
longer (within 30~min)---in many cases, this was even
longer than \tool's tuning time in \figref{fig:alg-total-time}. We
compared the number of assertions verified by our tailored recipes
(which do not exceed 5~min) and by the most precise recipe. For the 86
files that terminated within 5~min, our recipes prove 618 assertions,
whereas the most precise recipe proves 534. For the other 13 files,
our recipes prove 119 assertions, whereas the most precise recipe
proves 98.

Consequently, our (in theory) less precise and more efficient recipes
prove more assertions in files where the most precise recipe
terminates.  Possible explanations for this non-intuitive result
are: (1)~Polyhedra coefficients may overflow, in which case
the constraints are typically ignored by abstract interpreters, and
(2)~more precise domains with different widening operations may
result in less precise results~\cite{MonniauxLeGuen2012,
  AmatoRubino2018}.

\begin{figure}[b]
\centering
\scalebox{0.75}{
\begin{tikzpicture}
\begin{axis}[
   ybar stacked,
     symbolic x coords={1, 2, 3, 4, 5, 6, 7},
	 xtick=data,
	 xticklabels={\textsc{dw}, \textsc{ni}, \textsc{wt}, \textsc{as}, \textsc{b}, \textsc{d}, \textsc{o}},
     ylabel={\textbf{Percentage of mutated recipes}},
     legend cell align=left,
     axis y line*=none,
     axis x line*=bottom,
     width=10cm,
     height=5.5cm,
     ymin=0,
     ymax=125,
     ytick distance=50,
     bar width=0.65cm,
     area legend,
     legend style={legend pos=north east,font=\small, draw=none, legend columns=-1}
   ]

\addplot+[ybar, DarkOrange!70] plot coordinates {(1,81.2) (2,90) (3,89.5) (4,69.9) (5,37.9) (6,14.2) (7,95.7)};

\addplot+[ybar, Red!70] plot coordinates {(1,10) (2,0.8) (3,1.4) (4,16.7) (5,57.1) (6,84) (7,4.3)};

\addplot+[ybar, Green!70] plot coordinates {(1,8.7) (2,8.8) (3,8.9) (4,13.3) (5,4.9) (6,1.7) (7, 0)};

\legend{Equal, Negative, Positive}
\end{axis}
\end{tikzpicture}
}
\vspace{-0.5em}
\caption{Effect of different settings on the precision and performance
  of the abstract interpreter. (\textsc{dw}: \delays, \textsc{ni}:
  \narrowings, \textsc{wt}: \thresholds, \textsc{as}: array smashing,
  \textsc{b}: backward analysis, \textsc{d}: abstract domain,
  \textsc{o}: ingredient ordering).}
\label{fig:toggle-best-recipe}
\end{figure}
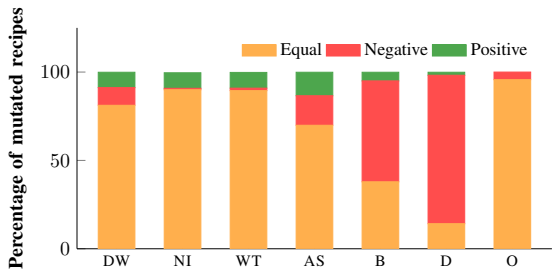

We also evaluated the optimality of tailored recipes by mutating
individual parts of the recipe and comparing to the original.
In particular, for each setting in
\tabref{tab:allowed-parameter-values}, we tried all possible values
and replaced each domain with all other comparable domains in the
poset of \figref{fig:lattice}.
For example, for a recipe including \texttt{zones}, we tried
\texttt{octagons}, \texttt{polyhedra}, and \texttt{intervals}.
In addition, we tried all possible orderings of the recipe
ingredients, which in theory could produce different results. We
observed whether these changes resulted in a difference in the
precision and performance of the analyzer.

\figref{fig:toggle-best-recipe} shows the results of this experiment,
broken down by setting. Equal (in orange) indicates that the mutated
recipe proves the same number of assertions within $\pm 5$ seconds of
the original. Positive (in green) indicates that it either proves more
assertions or the same number of assertions at least $5$ seconds
faster.  Negative (in red) indicates that the mutated recipe either
proves fewer assertions or the same number of assertions at least $5$
seconds slower.

The results show that, for our benchmarks, mutating the recipe found
by \tool rarely led to an improvement. In particular, at least
93\% of all mutated recipes were either equal to or worse than the
original recipe. In the majority of these cases, mutated recipes are
equally good. This indicates that there are many optimal or
close-to-optimal solutions and that \tool is able to find one
of them. \\

\fbox{
  \begin{minipage}{0.87\linewidth}
    {\bf RQ2 takeaway:}
    As compared to the most precise recipe, \tool verified
    more assertions across benchmarks where the most precise
    recipe terminated. Furthermore, mutating recipes found by \tool
    resulted in improvement only for less than 7\% of recipes.
  \end{minipage}
} \\

\textbf{RQ3: How diverse are the tailored recipes?}
To motivate the need for optimization, we must show that tailored
recipes are sufficiently diverse such that they could not be replaced
by a well-crafted default recipe. To better understand the
characteristics of tailored recipes, we manually inspected all recipes
generated by \tool.

\tool generated recipes of length greater than 1 for 61 files. Out of
these, 37 are of length 2 and 24 of length 3.  For 77\% of generated
recipes, \delays is not set to the default value of $1$. Additionally,
55\% of the ingredients enable array smashing, and 32\% enable
backward analysis.


\begin{filecontents*}{data/most-used-domains.csv}
Name,Quantity
"\texttt{bool}",18.2
"\texttt{ric}",12.3
"\texttt{term(int)}",11.8
"\texttt{polyhedra}",11.3
"\texttt{term(disInt)}",10.8
"\texttt{octagons}",10.8
"\texttt{boxes}",9.3
"\texttt{zones}",5.9
"\texttt{disInt}",4.9
"\texttt{intervals}",4
\end{filecontents*}
\DTLloaddb{data}{data/most-used-domains.csv}
\begin{filecontents*}{data/most-used-domains-bufferoverflow.csv}
Name,Quantity
"\texttt{bool}",58.9
"\texttt{boxes}",10.5
"\texttt{int}",6.8
"\texttt{polyhedra}",6.2
"\texttt{octagons}",4.3
"\texttt{term(int)}",3.7
"\texttt{disInt}",3.1
"\texttt{term(disInt)}",2.4
"\texttt{zones}",2.4
"\texttt{ric}",1.8
\end{filecontents*}
\DTLloaddb{dataBO}{data/most-used-domains-bufferoverflow.csv}
\begin{filecontents*}{data/most-used-domains-divbyzero.csv}
Name,Quantity
"bool",40.3
"term(disInt)",26
"intervals",15.1
"zones",8.4 
"ric",4.2
"polyhedra",2.5
"boxes",2.5
"oct",0.8

\end{filecontents*}
\DTLloaddb{dataDBZ}{data/most-used-domains-divbyzero.csv}
\begin{filecontents*}{data/most-used-domains-intoverflow.csv}
Name,Quantity
"\texttt{bool}",17.4
"\texttt{boxes}",13.1
"\texttt{polyhedra}",12.6
"\texttt{octagons}",9.7
"\texttt{ric}",9.2
"\texttt{term(disInt)}",9.2
"\texttt{term(int)}",8.2
"\texttt{disInt}",7.2
"\texttt{zones}",6.7
"\texttt{intervals}",6.3
\end{filecontents*}
\DTLloaddb{dataIO}{data/most-used-domains-intoverflow.csv}

\begin{figure}[t]
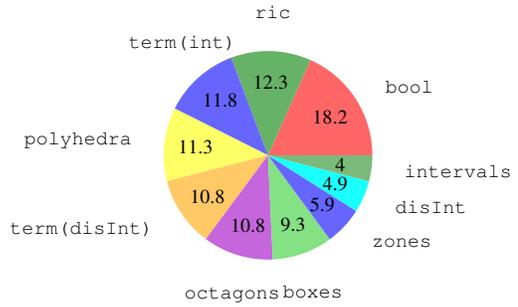

\centering
\scalebox{0.75}{
\DTLpiechart{variable=\quantity,outerlabel=\name,innerratio=0.7,radius=1.85cm}{data}{
  \name=Name,\quantity=Quantity}}
\vspace{-0.5em}
\caption{Occurrence of domains (in \%) in the best recipes
  found by \tool for all assertion types.}
\vspace{-0.5em}
\label{fig:most-used-domains}
\end{figure}

\begin{figure}[t]
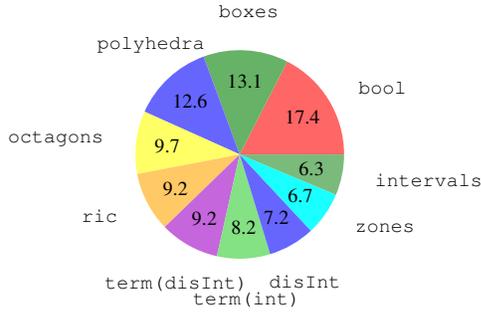

  \centering
  \scalebox{0.75}{
\DTLpiechart{variable=\quantity,outerlabel=\name,innerratio=0.7,radius=1.85cm}{dataIO}{
  \name=Name,\quantity=Quantity}}
\vspace{-0.5em}
\caption{Occurrence of domains (in \%) in the best recipes
  found by \tool for integer-overflow assertions.}
\vspace{-0.5em}
\label{fig:most-used-domains-IO}
\end{figure}

\figref{fig:most-used-domains} shows how often (in percentage) each
abstract domain occurs in a best recipe found by \tool. We
observe that all domains occur almost equally often, with $6$ of the $10$
domains occurring in between $9$\% and $13$\% of recipes.
The most common domain was \texttt{bool} at $18$\%, and the least
common was \texttt{intervals} at $4$\%.
We observed a similar distribution of domains
even when instrumenting the benchmarks with only one assertion
type, e.g., those that check for integer overflow (see
\figref{fig:most-used-domains-IO}).

We also inspected which domain combinations are frequently used in the
tailored recipes. One common pattern is combinations between
\texttt{bool} and numerical domains (18 occurrences).
%
%
Similarly, we observed 2 occurrences of \texttt{term(disInt)} together
with \texttt{zones}.
%
%
Interestingly, the less powerful variants of combining \texttt{disInt}
with \texttt{zones} (3 occurrences) and \texttt{term(int)} with
\texttt{zones} (6 occurrences) seem to be sufficient in many
cases. Finally, we observed 8 occurrences of \texttt{polyhedra} or
\texttt{octagons} with \texttt{boxes}, which are the most precise
convex and non-convex domains. Our approach is, thus, not only useful
for users, but also for designers of abstract interpreters by
potentially inspiring new domain combinations. \\

\fbox{
  \begin{minipage}{0.87\linewidth}
    {\bf RQ3 takeaway:} The diversity of tailored recipes prevents
    replacing them with a single default recipe. Over half of the
    tailored recipes contain more than one ingredient, and ingredients
    use a variety of domains and their settings.
  \end{minipage}
} \\

\textbf{RQ4: How resilient are the tailored recipes to code changes?}
We expect tailored recipes to be resilient to code changes, i.e., to
retain their optimality across several changes without requiring
re-tuning. We now evaluate if a recipe tailored for one code version
is also tailored for another, even when the two versions are 50
commits apart.

For this experiment, we took a random sample of 60 files from our
benchmarks and retrieved the 50 most recent commits per file. We only
sampled 60 out of 120 files as building these files for each commit is
quite time consuming---it can take up to a couple of days.
We instrumented each file version with the four assertion types
described in \secref{subsec:Benchmarks}. It should be noted that, for
some files, we retrieved fewer than 50 versions either because there
were fewer than 50 total commits or our build procedure for the
project failed on older commits. This is also why we did not run this
experiment for over 50 commits.

We analyzed each file version with the best recipe, $R_o$, found by
\tool for the oldest file version. We compared this recipe with new
best recipes, $R_n$, that were generated by \tool when run on each
subsequent file version. For this experiment, we used a 5-min timeout
and 40 iterations.

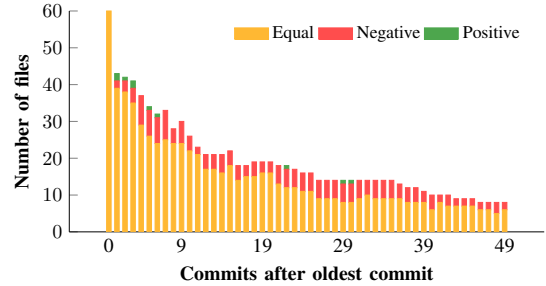
\begin{figure}[t]
\centering
\scalebox{0.75}{
\begin{tikzpicture}
\begin{axis}[
   ybar stacked,
   symbolic x coords={0, 1, 2, 3, 4, 5, 6, 7, 8, 9, 10, 11, 12,
     					13, 14, 15, 16, 17, 18, 19, 20,
                      21, 22, 23, 24, 25, 26, 27, 28, 29, 30,
                      31, 32, 33, 34, 35, 36, 37, 38, 39, 40, 41,
                     42, 43, 44, 45, 46, 47, 48, 49},
	 xtick=data,
	 xticklabels={0,,,,,,,,,9,,,,,,,,,,19,,,,,,,,,,29,,,,,,,,,,39,,,,,,,,,,49},
     ylabel={\textbf{Number of files}},
     xlabel={\textbf{Commits after oldest commit}},
     legend cell align=left,
     axis y line*=none,
     axis x line*=bottom,
     width=10cm,
     height=5.5cm,
     ymin=0,
     ymax=60,
     ytick distance=10,
     bar width=0.08cm,
     area legend,
     legend style={legend pos=north east,font=\small, draw=none, legend columns=-1}
   ]

  \addplot+[ybar, Orange!80] plot coordinates {
    (0, 60) (1, 39) (2, 38) (3, 35) (4, 29) (5, 26) (6, 24) (7, 25) (8, 24) (9, 24) (10, 22) (11, 21) (12, 17) (13, 17) (14, 16) (15, 18) (16, 14) (17, 15) (18, 15) (19, 16) (20, 16) (21, 13) (22, 12) (23, 12) (24, 11) (25, 11) (26, 9) (27, 9) (28, 9) (29, 8) (30, 8) (31, 9) (32, 10) (33, 9) (34, 9) (35, 9) (36, 9) (37, 8) (38, 8) (39, 8) (40, 6) (41, 8) (42, 7) (43, 7) (44, 7) (45, 7) (46, 6) (47, 6) (48, 5) (49, 6)
  };
  \addplot+[ybar, Red!70] plot coordinates {
    (0, 0) (1, 2) (2, 3) (3, 4) (4, 8) (5, 7) (6, 7) (7, 8) (8, 4) (9, 6) (10, 4) (11, 2) (12, 4) (13, 4) (14, 5) (15, 4) (16, 4) (17, 3) (18, 4) (19, 3) (20, 3) (21, 5) (22, 5) (23, 5) (24, 5) (25, 5) (26, 5) (27, 5) (28, 5) (29, 5) (30, 5) (31, 5) (32, 4) (33, 5) (34, 5) (35, 5) (36, 4) (37, 4) (38, 4) (39, 3) (40, 4) (41, 2) (42, 3) (43, 2) (44, 2) (45, 2) (46, 2) (47, 2) (48, 3) (49, 2)
  };
  \addplot+[ybar, Green!70] plot coordinates {
    (0, 0) (1, 2) (2, 1) (3, 2) (4, 0) (5, 1) (6, 1) (7, 0) (8, 0) (9, 0) (10, 0) (11, 0) (12, 0) (13, 0) (14, 0) (15, 0) (16, 0) (17, 0) (18, 0) (19, 0) (20, 0) (21, 0) (22, 1) (23, 0) (24, 0) (25, 0) (26, 0) (27, 0) (28, 0) (29, 1) (30, 1) (31, 0) (32, 0) (33, 0) (34, 0) (35, 0) (36, 0) (37, 0) (38, 0) (39, 0) (40, 0) (41, 0) (42, 0) (43, 0) (44, 0) (45, 0) (46, 0) (47, 0) (48, 0) (49, 0)
  };
\legend{Equal, Negative, Positive}
\end{axis}
\end{tikzpicture}
}
\vspace{-0.5em}
\caption{Difference in the safe assertions across commits.}
\label{fig:commits-different-seeds}
\vspace{-0.5em}
\end{figure}

\begin{figure}[t]
\centering
\scalebox{0.75}{
\begin{tikzpicture}
  \pgfplotsset{
      scale only axis,
  }

  \begin{axis}[
      xlabel=\vspace{0.5em}\textbf{LOC difference from oldest commit},
      ylabel style={align=center}, ylabel=\textbf{Average difference in safe assertions}\\\textbf{from oldest commit},
  ]

    \pgfplotstableread[col sep = comma]{data/loc-diff-forward-abs.csv}\scatterplot
  \addplot[scatter, only marks] table [x=xax, y=yax, col sep=comma] {\scatterplot};

  \end{axis}
\end{tikzpicture}
}
\vspace{-0.5em}
\caption{Average difference in the number of safe assertions per
  change in lines of code from oldest commit.}
\label{fig:commits-loc-different-seeds}
\vspace{-0.5em}
\end{figure}
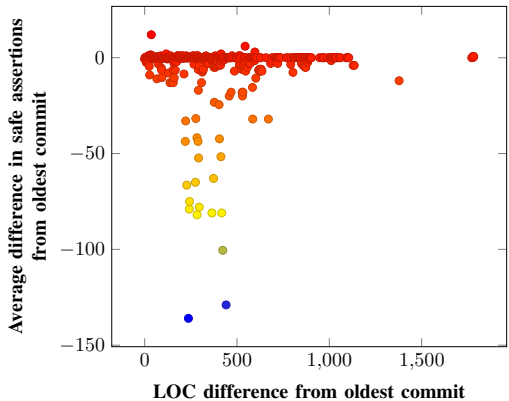

Note that, when running \tool with the same optimization algorithm and
random seed, it explores the same recipes. It is, therefore, very
likely that recipe $R_o$ for the oldest commit is also the best for
other file versions since we only explore 40 different recipes. To
avoid any such bias, we performed this experiment by seeding \tool
with a different random seed for each commit. The results are shown in
\figsref{fig:commits-different-seeds} and
\ref{fig:commits-loc-different-seeds}.

In \figref{fig:commits-different-seeds}, we give a bar chart comparing
the number of files per commit that have a positive, equal, and
negative difference in the number of verified assertions, where commit
$0$ is the oldest commit and $49$ the newest.  An equal difference
(in orange) means that recipe $R_o$ for the oldest commit proves the
same number of assertions in the current file version, $f_n$, as
recipe $R_n$ found by running \tool on $f_n$. To be more
precise, we consider the two recipes to be equal if they differ by at
most 1 verified assertion or 1\% of verified assertions since such a
small change in the number of safe assertions seems acceptable in
practice (especially given that the total number of assertions may
change across commits). A positive difference (in green) means that
$R_o$ achieves better verification results than $R_n$, that is, $R_o$
proves more assertions safe (over 1 assertion or 1\% of the assertions
that $R_n$ proves). Analogously, a negative difference (in red) means
that $R_o$ proves fewer assertions.  We do not consider time here
because none of the recipes timed out when applied on any file
version.

Note that the number of files decreases for newer commits.
This is because not all files go forward by 50 commits, and
even if they do, not all file versions build. However, in a few
instances, the number of files increases going forward in time. This
happens for files that change names, and later, change back, which we
do not catch.

For the vast majority of files, using recipe
$R_o$ (found for the oldest commit) is as effective as using $R_n$
(found for the current commit). The difference in safe assertions is
negative for less than a quarter of the files tested, with the average
negative difference among these files being around 22\% (i.e., $R_o$
proved 22\% fewer assertions than $R_n$ in these files). On the
remaining three quarters of the files tested however, $R_o$ proves at
least as many assertions as $R_n$, and thus, $R_o$ tends to be tailored
across code versions.

\figref{fig:commits-loc-different-seeds} shows the average difference
in the number of verified assertions per change in lines of code from
the oldest commit. Note that a positive (resp. negative) difference
represents that $R_o$ (resp. $R_n$) proves more assertions. The plot
clearly shows that for most files, regardless of lines changed, $R_o$
and $R_n$ are equally effective. Regarding the outliers shown in the
figure, we noticed that all commits with a difference of 50 safe
assertions or more modify one particular file from the \textsc{git}
project. In this case, $R_o$ is not as effective because the widening
and narrowing settings have very low values. \\

\fbox{
  \begin{minipage}{0.87\linewidth}
    {\bf RQ4 takeaway:} For over $75$\% of files, \tool's recipe for a
    previous commit (from up to 50 commits previous) remains tailored
    for future versions of the file, indicating the resilience of
    tailored recipes across code changes.
  \end{minipage}
} \\

\subsection{Threats to Validity}
\label{subsec:Threats}

We have identified the following threats to the validity of our
experiments.

\textbf{Benchmark selection.} Our results may not generalize to other
benchmarks. However, we selected popular GitHub projects from
different application domains (see \tabref{tab:projects}).
Hence, we believe that our benchmark selection mitigates this threat
and increases generalizability of our findings.

\textbf{Abstract interpreter and recipe settings.} For our
experiments, we only used a single abstract interpreter, \crab, which
however is a mature tool and actively supported.
The selection of recipe settings was, of course, influenced by the
available settings in \crab. Nevertheless, \crab implements the
generic architecture of \figref{fig:interpreter}, used by most
abstract interpreters, such as those mentioned at the beginning of
\secref{sec:abstract-interpreters}. We, therefore, expect our approach
to generalize to such analyzers.

\textbf{Optimization algorithms.} We considered four optimization
algorithms, but in \secref{subsec:Algorithms}, we explain why these
are suitable for our application domain. Moreover, \tool is
configurable with respect to the optimization algorithm.

\textbf{Assertion types.} Our results are based on four types of
assertions. However, these cover a broad spectrum of runtime errors
that are commonly checked by static analyzers.

\section{Related Work}
\label{sec:RelatedWork}

The impact of different abstract-interpretation configurations has
been previously evaluated~\cite{WeiMardziel2018} for Java programs and
partially inspired this work. To the best of our knowledge, we are the
first to propose tailoring static analyzers to custom usage scenarios
using optimization. However, optimization is a widely used technique
in many engineering disciplines. In the following, we focus on its use
in program analysis.


Optimization has been successfully applied to a number of
program-analysis problems, such as automated testing~\cite{FuSu2016,
FuSu2017}, invariant inference~\cite{SharmaAiken2014}, and compiler
optimizations~\cite{SchkufzaSharma2013}.

Many machine-learning techniques also rely on optimization, for
instance of loss functions in neural networks. Recently, researchers
have started to explore the direction of enriching program analyses
with machine-learning techniques, for example, to automatically learn
analysis
heuristics~\cite{HeoOh2016,JeongJeon2017,RaychevVechev2019,SinghPueschel2018-FastAnalysis}.
A particularly relevant body of work is on adaptive program
analysis~\cite{HeoOh2019,HeoOh2018,HeoOh2017}, where existing code is
analyzed to learn heuristics that trade soundness for precision or
that coarsen the analysis abstractions to improve memory consumption.

More specifically, adaptive program analysis poses different
static-analysis problems as machine-learning problems and relies on
Bayesian optimization to solve them, e.g., the problem of selectively
applying unsoundness to different program components (e.g., different
loops in the program)~\cite{HeoOh2017}. The main insight is that
program components (e.g., loops) that produce false positives are
alike, predictable, and share common properties. After learning to
identify such components for existing code, this technique suggests
components in unseen code that should be analyzed unsoundly.

In contrast, \tool currently does not adjust soundness of the
analysis. However, this would also be possible if the analyzer
provided the corresponding configurations. More importantly, adaptive
analysis focuses on learning analysis heuristics based on existing
code in order to generalize to arbitrary, unseen code. \tool, on the
other hand, aims to tune the analyzer configuration to a custom usage
scenario, including a particular program under analysis. In addition,
the custom usage scenario imposes user-specific resource constraints,
for instance by limiting the time according to a phase of the
software-engineering life cycle. As we show in our experiments, the
tuned configuration remains tailored to several versions of the
analyzed program. In fact, it outperforms configurations that are
meant to generalize to arbitrary programs, such as the default
recipe.




\section{Conclusion}
\label{sec:Conclusion}

In this paper, we have proposed a technique and framework that tailors
a generic abstract interpreter to custom usage scenarios. We
instantiated our framework with a mature abstract interpreter to
perform an extensive evaluation on real-world benchmarks.
%
%
Our experiments show that the configurations generated by \tool are vastly better than the
default options, vary significantly depending on the code under analysis, and
most remain tailored to several subsequent code versions.
%
In the future, we plan to explore the challenges that an
inter-procedural analysis would pose, for instance, by using a
different recipe for computing a summary of each function or each
calling context.



\bibliography{bibliography}

\end{document}